# The study of the photon structure function at the ILC energy range

Beata Krupa[a*], Tomasz Wojtoń[a], Leszek Zawiejski[a]

[a]*Institute of Nuclear Physics PAN, Cracow, Poland*

**Abstract**

At the future $e^+e^-$ linear collider ILC/CLIC it will be able to measure the photon structure functions in a wider range of kinematic variables $x$ and $Q^2$ in comparison to the previous results of experiments at LEP. The classical way to measure the photon structure functions is the study of $e^+e^- \to e^+e^- \gamma\gamma \to e^+e^- X$ process, where X is the leptonic or hadronic final state. For the study of the QED and hadronic photon structure functions the simulations of two-photon processes were performed at the ILC center-of-mass energy equal to 500 GeV using the PYTHIA and the ILCSoft package. The analysis used information related to the forward detectors, tracking detectors and calorimeters which are parts of the ILD detector concept.

*Keywords*: two-photon interactions, photon structure function

**1. Motivation and theoretical framework**

The two-photon processes provide a comprehensive laboratory for exploring virtually every aspect of the Standard Model and its extensions. They serve as the prototypes of collisions of other gauge bosons, allowing us to test the electroweak theory in photon-photon annihilation and to provide a good testing ground for studying the predictions of the quantum chromodynamics (QCD). These processes could be also the source of the production of supersymmetric squark and slepton pairs. In particular, the two-photon process in which one of the virtual photon is very far off-shell (large virtuality), while the other is close to the mass-shell (small virtuality), can be viewed as deep inelastic electron scattering off the photon [1]. This process of deep inelastic scattering is usually used to investigate the photon structure functions [2], which are analogous to the nucleon structure functions.

The first measurement of the photon structure function has been performed using the detector PLUTO at the DESY storage ring PETRA (1981) [3]. Following this pioneering work, many experiments have been performed at all high energy $e^+e^-$ and ep storage rings [4], where the lepton beams serve as a source of high energy photons. The classic way to investigate the structure of the photon at $e^+e^-$ colliders is the study of the process: $e^+e^- \to e^+e^- X$, proceeding via the interaction of two photons, which can be either quasi-real or virtual [2]. The incoming leptons radiate photons producing a hadronic or leptonic final state X. The kinematic of these interactions is illustrated in Figure 1, which also includes the definitions of the photons' virtualities ($Q^2$, $P^2$) and the invariant mass squared of $\gamma\gamma$ system ($W^2$). The polar angles at which the electrons are scattered are measured with respect to the direction of the beam electrons and depend on the virtualities of the photons.

From the experimental point of view the following three event classes can be distinguished. In the case where none of the scattered beam electrons can be observed in the detector (anti-tagged events), one can study the structure of a quasi-real photon in terms of total cross-sections, jet production and heavy quark production. If both electrons are observed (double-tagged events), the dynamics of highly virtual photon collisions is probed. And if only one electron is detected (single-tagged events), the process can be described as deep inelastic electron scattering off a quasi-real photon. These events can be studied to measure both QED and hadronic photon structure functions [5]. In this paper we focus on the study of the quasi-real photons, therefore the single-tagged events will be further analysed.

---

[*] e-mail: beata.krupa@ifj.edu.pl

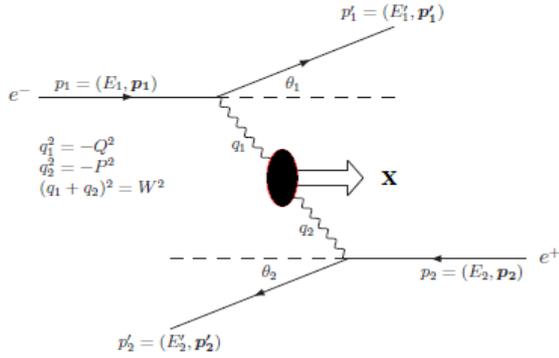

Fig. 1. Kinematics of the two photon process [4].

The usual dimensionless kinematical variables of deep inelastic scattering are the fraction of parton momentum with respect to the target photon ($x$) and the energy lost by the inelastically scattered electrons ($y$), which are defined as:

$$x = \frac{Q^2}{2q_1q_2}, \quad y = \frac{q_1q_2}{p_1q_2}.$$

Experimentally, the kinematical variables $Q^2$, $x$ and $y$ are obtained from the four-vectors of the tagged electrons and the hadronic or leptonic final state using the information about the polar angle at which the tagged electron is scattered (discussed in detail in [6]) via:

$$Q^2 = 4E_bE\sin^2(\theta/2)$$
$$x = \frac{Q^2}{Q^2+W^2+P^2}$$
$$y = 1 - \frac{E}{E_b}\cos^2(\theta/2)$$

where $E_b$ is the energy of the beam electrons, $E$ and $\theta$ refer to the energy and the polar angle of the scattered electrons. The differential cross section can be written in terms of $x$ and $y$ variables [7]:

$$\frac{d^2\sigma}{dQ^2dx} = \frac{2\pi\alpha^2}{xQ^4}([1+(1-y)^2]F_2^\gamma(x,Q^2) - y^2F_L^\gamma(x,Q^2)),$$

where $F_2^\gamma(x,Q^2)$ and $F_L^\gamma(x,Q^2)$ are the structure functions.

In spite of many studies of the photon structure, still it is needed to bring our understanding of the photon to the same level as HERA has achieved for the proton. This will offer new insights into QCD.

Since the beam energy at the future linear collider ILC/CLIC will be higher, it is expected that it will be possible to measure the evolution of the photon structure function in a wider range of $Q^2$ and $x$ variable. It would be interesting to study the structure function for highly virtual photons, because the interaction of two virtual photons is the so-called 'golden' process to study the parton dynamics (DGLAP and/or BFKL) [8]. For this purpose, the ability to tag both scattered electrons (double-tagged events) is needed. It would allow us also to determine the invariant mass squared of $\gamma\gamma$ system $W^2$ independently of the hadronic final state and thus to increase the precision of the measurement of the photon structure function. Moreover, a new light on the photon structure would be shed by spin-dependent structure functions, which have not been measured so far. This would be possible in the polarized $e^+e^-$ collisions at the future linear collider.

## 2. Expected values of kinematic variables

As noted above, since the beam energy at the future linear collider ILC/CLIC is planned to be higher than in the previously performed experiments, it is expected that the lower limit of $x$ variable will be moved towards lower values. Extending the range of available values of $Q^2$ is also anticipated, which was confirmed by the Monte Carlo simulations carried out with the use of PYTHIA 6.4 [9]. The results of these simulations in the case of tagging the scattered electrons in deep inelastic $e\gamma$ scattering at the LumiCal and BeamCal detectors are presented in Figure 2. The blue markers are related to the generated events at the ILC centre-of-mass energy equal to 500 GeV. And for comparison, the red markers indicate the generated events at the CLIC centre-of-mass energy equal to 3 TeV.

Furthermore, Figure 3 and Figure 4 show the histograms which are the predictions for the distributions of other kinematic variables: $x$ and $y$ respectively in the case of single-tagged events at the ILC centre-of-mass energy equal to 500 GeV with scattered electrons detected only at the LumiCal detector. By virtue of the weak dependence of the *x* distribution on the target photon virtuality ($P^2$), the zero value of $P^2$ can be assumed (as in the previous experiments, e.g. at LEP [2]).

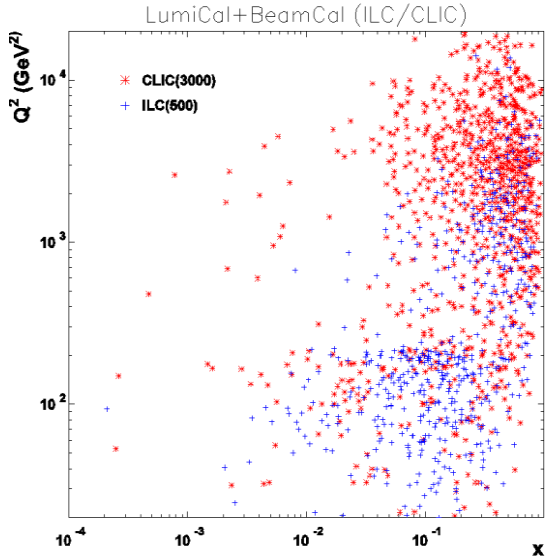

Fig. 2. Kinematical plane (*x*, $Q^2$) with simulated single-tagged events for the case of detecting scattered electrons at the LumiCal and BeamCal detectors.

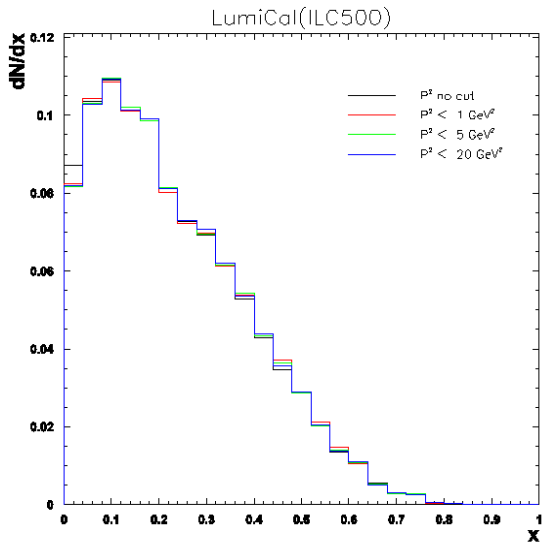

Fig. 3. Distribution of *x* variable. The histograms are the predictions from the PYTHIA for the case of single-tagged events at the ILC centre-of-mass energy equal to 500 GeV with scattered electrons detected only at the LumiCal detector. The lines of different colours refer to various cuts on the virtuality on the target photon ($P^2$).

Moreover, the obtained mean value of *y* variable (Fig. 4) is less than 0.12, thus the $F_L^\gamma$ term in the expression for the differential cross section can be neglected. Therefore, by measuring the differential cross section one can determine $F_2^\gamma$ function.

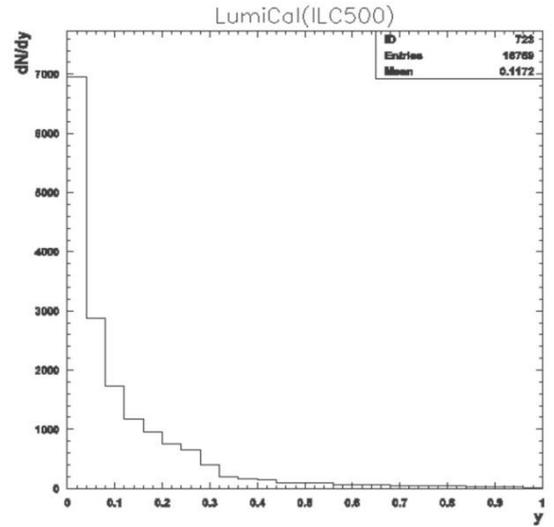

Fig. 4. Distribution of *y* variable. The histogram is the prediction from the PYTHIA for the case of single-tagged events at the ILC centre-of-mass energy equal to 500 GeV with scattered electrons detected only at the LumiCal detector.

### 3. Event selection

The study of the possibility to measure the $F_2^\gamma$ structure function of the quasi-real photon is based on the Monte Carlo simulations of single-tagged events with a leptonic (QED structure function) or hadronic (hadronic structure function) final state in deep inelastic $e\gamma$ scattering regime. At first it is assumed that scattered electrons are detected only at the LumiCal detector. Such events can be selected with the following set of cuts.

Firstly, an electron candidate should be observed in the LumiCal detector with energy $E_{tag} > 0.7\ E_b$ and polar angle in the range of the angular acceptance of this detector, i.e. $31 < \theta < 78$ mrad. The angle $\theta$ is measured with respect to the original beam direction. Furthermore, there must be no deposited energy with value $E_a > 0.2\ E_b$ in the detector on the opposite side. This is an anti-tag cut which is applied for possible electron candidates in the hemisphere opposite to the tagged electron in order to guarantee the low virtuality of the quasi-real photon.

Secondly, in the case of the determination of hadronic structure function at least three tracks originating from a hadronic final state have to be present. When the QED structure function is studied, the events with muons in the final state are selected since this process gives the clearest measurement. This is because for $e^+e^-$ final state the number of different Feynman diagrams contributing to this process makes the analysis more difficult. On the other hand, for $\tau^+\tau^-$ final state, the final state can be only identified by detecting the products of $\tau$ decays, which is also more difficult. Muon pairs are detected in muon detectors.

Another requirement is that the visible invariant mass $W_{vis}$ should be in the range of 3 GeV $< W_{vis} <$ 0.6 $E_b$. The upper limit should reduce the expected background from annihilation events.

## 4. The first results for the $F_2^\gamma$ structure function

Monte Carlo simulations indicate that information from the LumiCal detector can be used to study the photon structure function. However, in order to extend the range of $x$ and $Q^2$ variables it is necessary to use also information from other detectors, such as BeamCal and ECAL. The results of PYTHIA 6.4 Monte Carlo studies are presented in Figures 5 and 6.

The errors marked on these plots are statistical only. Of course, the systematic effects should be estimated and the possible background should be considered. This will be, besides the use of the reconstructed variables, the next step of the analysis. It is also intended to compare the PYTHIA generator level results with results from other Monte Carlo generators, such as WHIZARD, HERWIG as well as those used at the LEP experiments after their adaptation to the ILC/CLIC conditions (e.g. PHOJET, TWOGAM, BDK).

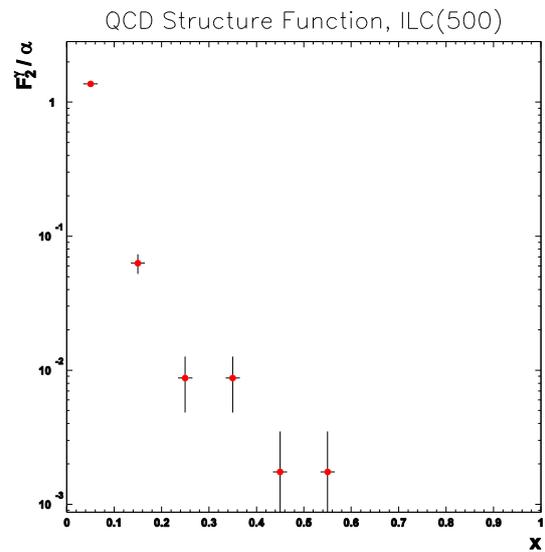

Fig. 6. The hadronic photon structure function divided by the fine structure constant as a function of *x* variable for the mean value of $Q^2$ equal to 119 GeV$^2$. The errors are statistical only.

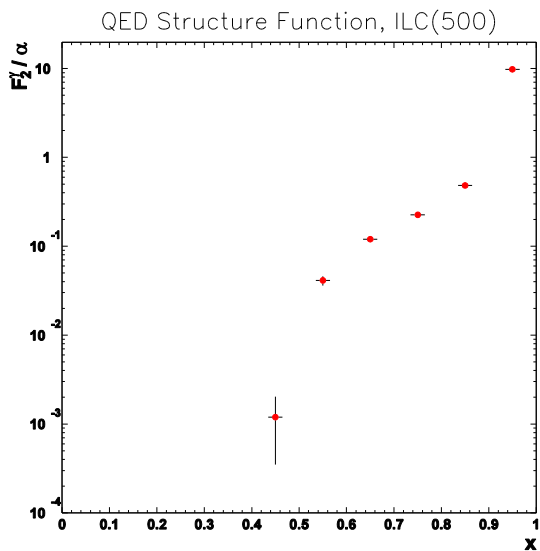

Fig. 5. The QED photon structure function divided by the fine structure constant as a function of *x* variable for the mean value of $Q^2$ equal to 119 GeV$^2$. The errors are statistical only.

**References**

[1] T. F. Walsh, Phys. Lett. 36B (1971) 121.
[2] M. Przybycień, *Study of the photon structure at LEP*, Habilitation Thesis, University of Science and Technology, Cracow, 2003.
[3] PLUTO Collaboration, Ch. Berger et al., Phys. Lett. **B107** (1981) 168.
[4] Ch. Berger, *Photon structure function revisited*, arXiv:1404.3551v1 [hep-ph].
[5] R. Nisius, Physics Reports 332 (2000) 165.
[6] V. M. Budnev et al., Physics Reports 15 (1975) 181.
[7] R. M. Godbole, *Photon structure function*, arXiv:hep-ph/9602428.
[8] G. P. Salam, Acta Phys. Polon. B30 (1999) 3679
[9] T. Sjöstrand, S. Mrenna, P. Skands, *PYTHIA 6.4 – physics and manual*, IHEP 05 (2006) 026.